\documentclass[english,aps,prb,showpacs,superscriptaddress,floatfix]{revtex4}
\usepackage{ae,aecompl}
\usepackage[T1]{fontenc}
\usepackage{color}
\usepackage{textcomp}
\usepackage{amssymb}
\usepackage{graphicx}
\makeatletter
\@ifundefined{textcolor}{}
{%
 \definecolor{BLACK}{gray}{0}
 \definecolor{WHITE}{gray}{1}
 \definecolor{RED}{rgb}{1,0,0}
 \definecolor{GREEN}{rgb}{0,1,0}
 \definecolor{BLUE}{rgb}{0,0,1}
 \definecolor{CYAN}{cmyk}{1,0,0,0}
 \definecolor{MAGENTA}{cmyk}{0,1,0,0}
 \definecolor{YELLOW}{cmyk}{0,0,1,0}
 }

\makeatother

\begin{document}

\title{Emergence of ferromagnetism and Jahn-Teller distortion
in low $Cr-$substituted $LaMnO_{3}$ }

\author{Aline Y. Ramos}

\email{aline.ramos@grenoble.cnrs.fr}

\affiliation{Institut N\'eel, CNRS and UJF, BP 166, F-38042 Grenoble Cedex 9,
France}

\author{H\'elio C. N. Tolentino}

\affiliation{Institut N\'eel, CNRS and UJF, BP 166, F-38042 Grenoble Cedex 9,
France}

\author{M\'arcio M. Soares}

\affiliation{Institut N\'eel, CNRS and UJF, BP 166, F-38042 Grenoble Cedex 9,
France}

\author{St\'ephane Grenier}

\affiliation{Institut N\'eel, CNRS and UJF, BP 166, F-38042 Grenoble Cedex 9,
France}

\author{Oana Bun{\u{a}}u}

\affiliation{Institut N\'eel, CNRS and UJF, BP 166, F-38042 Grenoble Cedex 9,
France}

\author{Yves Joly}

\affiliation{Institut N\'eel, CNRS and UJF, BP 166, F-38042 Grenoble Cedex 9,
France}

\author{Fran\c cois Baudelet }

\affiliation{Synchrotron SOLEIL, L'Orme des Merisiers, Saint-Aubin, BP 48, 91192
Gif-sur-Yvette Cedex, France}

\author{Fabrice Wilhelm}

\affiliation{European Synchrotron Radiation Facility - ESRF, F-38043 Grenoble,
France }

\author{Andrei Rogalev}

\affiliation{European Synchrotron Radiation Facility - ESRF, F-38043 Grenoble,
France }

\author{Raquel A. Souza}

\affiliation{Laborat\'orio Nacional de Luz S\'incrotron - P.O. Box 6192, 13084-971,
Campinas, S$\tilde{a}$o Paulo, Brazil}

\author{Narcizo M. Souza-Neto}

\affiliation{Laborat\'orio Nacional de Luz S\'incrotron - P.O. Box 6192, 13084-971,
Campinas, S$\tilde{a}$o Paulo, Brazil}

\author{Olivier Proux }

\affiliation{Observatoire des Sciences de l'Univers - OSUG- Grenoble, F-38051
Grenoble, France}

\author{Denis Testemale}

\affiliation{Institut N\'eel, CNRS and UJF, BP 166, F-38042 Grenoble Cedex 9,
France}

\author{Alberto Caneiro}

\affiliation{Centro At\'omico Bariloche, CNEA and Universidad Nacional de Cuyo,
8400 S.C. de Bariloche, Argentina}
\begin{abstract}
The emergence of a ferromagnetic component in $LaMnO_{3}$ with low
Cr-for-Mn substitution has been studied by x-ray absorption spectroscopy
and x-ray magnetic circular dichroism at the Mn and Cr K edges. The
local magnetic moment strength for the Mn and Cr are proportional
to each other and follows the macroscopic magnetization. The net ferromagnetic
components of $Cr^{3+}$ and $Mn^{3+}$ are found antiferromagnetically
coupled. Unlike hole doping by La site substitution, the inclusion
of $Cr^{3+}$ ions up to x = 0.15 does not decrease the Jahn-Teller
(JT) distortion and consequently does not significantly affect the
orbital ordering. This demonstrates that the emergence of the ferromagnetism
is not related to JT weakening and likely arises from a complex orbital
mixing. 
\end{abstract}


\maketitle
Doped manganites have been extensively studied owing to their colossal
magneto resistance property and its potential applications to magnetic
devices. The substitution at the Mn site (B site in the $ABO_{3}$
perovskite formula) of various transitional metal elements dramatically
modifies the magnetic and electronic properties. The variations of
the magnetic and electronic behaviors are associated with the dissimilar
3d electron configuration of the substitutions. Cr doping at the Mn
sites has attracted special attention because $Cr^{3+}$ ions have
the same electronic configuration $t_{2g}^{3}e_{g}^{0}$ as $Mn^{4+}$.
As the $Cr^{3+}$~ionic radius (0.615~{\AA}) is much closer to that
of $Mn^{3+}$~high spin (0.645 {\AA}) than $Mn^{4+}$ ~radius (0.530~{\AA}),
low substitutions of $Cr^{3+}$~result in changing the $Mn^{3+}$~density
without large distortion of the crystal cell\cite{Morales2005PRB72}.
These features make the partial substitution of $Mn$ by $Cr$ in
$LaMnO_{3.00+\delta}$ an interesting system for studying the close
relationship between Jahn Teller (JT) distortion, double exchange
(DE) and superexchange (SE) interactions and orbital degeneracy. Since
the first studies of the series $LaMn_{1-x}Cr_{x}O_{3}$ in the 1950s\cite{BentsPRB1957,JonkerPhysica56}
it is known that upon Cr-doping the antiferromagnetic (AFM) Mott insulator
$LaMnO_{3}$ develops a ferromagnetic (FM) component, but the character
of the $Mn^{3+}-Cr^{3+}$ interaction has been under debate for a
long time \cite{Morales2005PRB72,Ganguly_PhysicaB00,GundakaramJSSC1996,Qu_PRB06,Sun2001PRB63,Farell_NewJPhys04,Cabeza_JPCM99,Barnabe_APL97,BentsPRB1957,CapognaPRB08,Ono_JMMM01,Toulemonde_EPL98,Terashita-PRB12,Cezar-JSR10,DeisenhoferPRB66}.
A $Mn^{3+}$-$Cr^{3+}$ SE FM coupling may account for the increasing
Curie temperature and spontaneous magnetization with increasing Cr
doping \cite{Ganguly_PhysicaB00,GundakaramJSSC1996,Sun2001PRB63,Qu_PRB06}.
On the other hand several groups proposed DE interaction between $Mn^{3+}$
and $Cr^{3+}$~ \cite{Sun2001PRB63,Morales2005PRB72}. Cr dopants
can also be viewed as a quenched random field \cite{Farell_NewJPhys04,Cabeza_JPCM99}.
However direct x-ray magnetic circular dichroism experiments suggest
that the exchange interaction between $Mn^{3+}$ and $Cr^{3+}$ leads
to a net local antiparallel coupling \cite{Toulemonde_EPL98,Ono_JMMM01,Cezar-JSR10,Terashita-PRB12},
regardless of the magnetic state nature of the manganite compound.
This scenario is supported by spin resonance investigations \cite{DeisenhoferPRB66}. 

As far as we know, unlike A site substitution, the local distortion
in Cr-substituted systems has not been the object of a specific study.
The usual view is that, in a similar way as the formal introduction
of $Mn^{4+}$by A site doping and/or oxygen overstoichiometry, B site
$Cr^{3+}$ substitution weakens the cooperative JT distortion, and
consequently affects the orbital order and favours the onset of FM
ordering. In contrast to this scenario where local site symmetrization
drives the magnetic behaviour, Zhou et al. \cite{Zhou_PRB08} recently
proposed that, for Ga-substituted doped manganites, local site distortions
could bias the orbital ordering so as to make orbital mixing responsible
for the 3D ferromagnetism. In order to clarify the spin-spin interactions
and ascertain the relative importance of removing a JT ion and of
hole doping, we performed coupled magnetic and local structural studies
by x-ray absorption near edge spectroscopy (XANES) and x-ray magnetic
circular dichroism (XMCD) at the transition metals K-edge, in $LaMn_{1-x}Cr_{x}O_{3.00}$
with low substitution of $Mn^{3+}$ by $Cr^{3+}$. We chose to study
specifically the emergence of the FM component in $Mn^{4+}$-free
compounds, with limited crystallographic distortions with respect
to $LaMnO_{3}$. This serie of compounds presents the double advantage
of avoiding additional complexity due to $Mn^{3+}-Mn^{4+}$ and $Mn^{4+}-Cr^{3+}$
 magnetic interactions, and the mixture of $Mn^{3+}$ and $Mn^{4+}$
local sites in mixed-valence compounds, and of starting from the parent
compound $LaMnO_{3.00}$ where the local structure is well characterized
\cite{Mitchell_PRB96,Souza-PRB04,MonesiPRB05,Ramos-EPL11}. The physics
studied in our paper concerns then the first steps in the substitution
of Mn atoms by Cr and is relatively simple because there is no $Mn^{4+}$ions
and $Cr{}^{3+}$-$Cr^{3+}$ interactions may be considered as negligible.

In this paper we show that the local magnetic moments at the Mn and
Cr atoms are proportional to the ferromagnetic component in the macroscopic
magnetization, and that their alignement is always antiparallel. We
found that Cr is assimilated in the Mn network and that the Cr environment
is essentially kept similar to that of Mn atoms, with no significant
reduction in the distortion, even if the $Cr^{3+}$ is not a Jahn-Teller
ion. This contradicts the usual guess\cite{Cabeza_JPCM99,Morales2005PRB72,Terashita-PRB12}
and indicates that JT weakening is not responsible for the emergence
of ferromagnetism, which should otherwise arise from orbital mixing. 

Polycrystalline powdered $LaMn_{1-x}Cr_{x}O_{3}$ with x= 0.05, 0.1
and 0.15 (corresponding to 5, 10 and 15\% of additional $e_{g}$ holes)\cite{Morales2004JAC,Morales2006PhysB}
and $LaMn_{0.90}Cr_{0.10}O_{3.04}$ (18\% $e_{g}$ holes, 8\% being
provided by oxygen over-stoichiometry - $Mn^{4+}$ ions)\cite{Morales2008JSSC}
were prepared by the liquid-mix method, using an experimental procedure
elaborated to enable an accurate control of the sample stoichiometry\cite{Morales2003JSSC}.
These compounds crystallize in the orthorhombic structure ($Pnma$).
They keep the $LaMnO_{3}$ A-AFM spin structure, ie FM ordering in
the ab planes and AFM between adjacents planes, but exhibit a significant
ferromagnetic component along the c axis \cite{Morales2004JAC,Morales2006PhysB}.
Room temperature (RT) XANES spectra at the Mn K edge (6539 eV) were
collected in the transmission mode at the Brazilian Synchrotron Light
Laboratory (LNLS) XAS1 beamline\cite{Tolentino-JSR01}. To check the
strict stoechiometry of our samples, a XANES reference of metal foil
was collected simultaneously. No edge shift is observed between the
$Mn^{4+}$-free samples and the $LaMnO_{3}$ standard, within the
precision of the experiments (0.1eV). On the contrary, for the non
stoichiometric $LaMn_{0.90}Cr_{0.10}O_{3.04}$ sample, we observe
an edge shift of about 0.6 eV  with respect to XANES
of $LaMnO_{3}$, in agreement with the increase in the formal valence.
RT XANES spectra at the Cr K edge (5989 eV) were collected at the
French CRG - BM30B beamline at the European Synchrotron Radiation
Facility (ESRF) using the fluorescence mode with a 30-element Ge solid
state detector \cite{Proux-JSR06}. The energy resolution, including
experimental resolution and core hole width, was about 1.5 eV at the
Cr K edge and 2 eV at the Mn K edge. The XANES spectra were normalized,
after background substraction, at about 200 eV above the edge for
the Mn edge and 50 eV for Cr edge. Coupled Mn and Cr K-edge XMCD measurements
were performed at the ID12 beamline at the ESRF, using the total fluorescence
yield mode with a Si photodiode detector and degree of circular polarization
of the monochromatic beam of 88\%. The samples were kept at 10 K and
magnetized by applying a 3 Tesla magnetic field. The temperature dependence
of the Mn XMCD was studied at the ODE dispersive beamline at the synchrotron
SOLEIL, using the transmission mode and a CCD detector \cite{BaudeletHPR11}.
The sample temperature was varied from 30 K to room temperature. XMCD
spectra were recorded by reversing a permanent magnet of 1.1 Tesla
magnetic field and maintaining the polarization of the light, with
a degree of circular polarization around 80\%. The XMCD signals resulting
from several scans were added and normalized to the absorption jump.
Slight differences between the signals collected in the transmission
(temperature dependent Mn K XMCD) and the fluorescence modes (coupled
10 K XMCD at Mn and Cr K edges) arise from differences in the amplitude
of the contribution of La edge\cite{Cezar-JSR10}. At the Cr K edge,
specially, the signal is superimposed to a large background contribution
of the La $L_{2}$ edge signal. This contribution is lower in the
fluorescence mode so that the main part of the Cr K edge XMCD data
were collected in this mode. After substraction of this contribution
the spectral shape is the same in both cases and the XMCD amplitude
is identical for a sample measured in the same conditions in the fluorescence
and in the transmission mode. 

Due to the selection rules in X-ray absorption spectroscopy, the K
edge transition originates from the core 1s state to the projected
np (mainly 4p) unoccupied density of states (DOS). X-ray absorption
probes the partial-DOS modified by the presence of the 1s core hole
that sorts out the 4p states around the Mn site from the ground state
band structure. The XANES signal at the Mn K edge (figure 1) is dominated
by a main line B, essentially related to the first coordination shell,
followed by structures (C and D) arising from multiple scattering
events. The pre-edge feature is formed by two peaks (P1, P2) and a
shoulder (A) associated with transition from Mn $1s$ levels to $4p$
empty levels\cite{Ramos-EPL11,Souza-PRB04,Bridges-PRB00}. XMCD at
the K edge of the transition metals probes the orbital polarization
of the conduction p-states, related to the spin polarization through
the spin\textendash{}orbit interaction \cite{Igarashi-PRB94} Consequently,
the XMCD at the K edge is, in its integral form, a measure of the
orbital magnetism of the 4p shell of the transition metal probed by
the x-rays. The delocalized 4p states are coupled by exchange-interaction
with the spin-polarized $3d\,$ band, which dominates the magnetic
properties of the system. Therefore the$\, K$~edge XMCD intensity
is proportional to the orbital magnetic moment of the~$3d$ bands
\cite{Guo-PRB98}. 

\begin{figure}
\includegraphics [bb=80bp 45bp 750bp 380bp,scale=1] {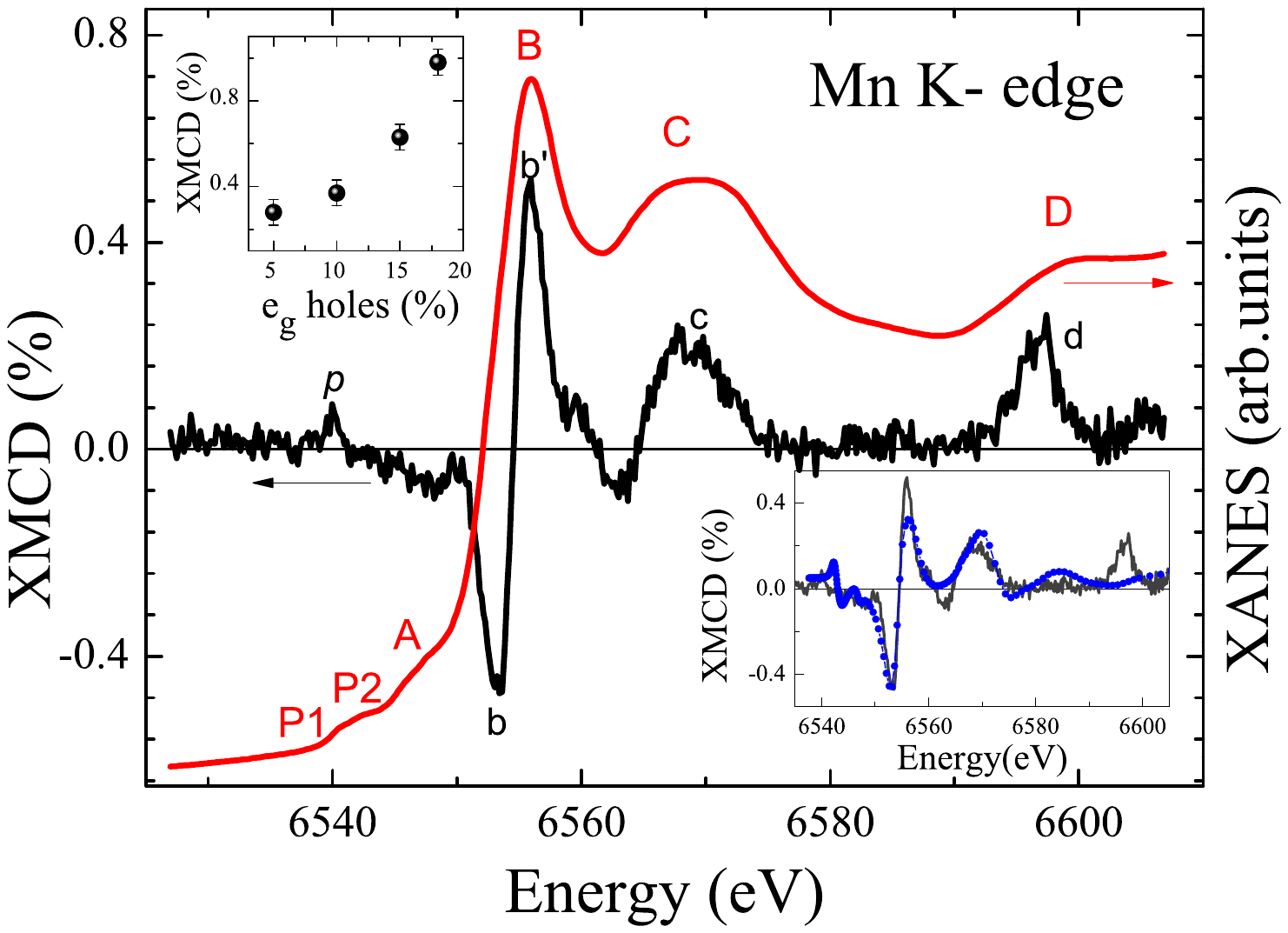}
\caption{XANES and XMCD (10 K, 3 Tesla) at the Mn K edge for $LaMn_{0.9}Cr_{0.1}O_{3.04}$.
The value peak b to peak b' is taken as a measure of the XMCD signal
intensity. Top inset: XMCD signal as a function of number of $e_{g}$
holes. Lower inset:\textit{\textcolor{black}{{} ab initio}} calculation
of the XMCD signal (dots) compared to experimental (line). }
\end{figure}

\textit{Ab initio} calculations of the XANES and XMCD features were
 performed using the FDMNES code \cite{Bunau-JPCM09}. The electronic
structure around the absorbing atoms is calculated using the multiple
scattering theory within the muffin-tin approximation, based on a
monoelectronic approach. The absorption is convoluted to a Lorentzian
with an energy dependent width, to take into account the core and
final state lifetimes and with a Gaussian to mimic the experimental
energy resolution. Calculations are performed for clusters built from
crystallographic data. In the Mn edge calculations we do not include
Cr atoms in the cluster. The Cr edge XANES are calculated considering
the only substitution of the Cr as central atom of the cluster. Cluster
of 33 atoms (radius 4.5 {\AA}) are large enough to reproduce the XMCD
features, while clusters up to 81 atoms (radius 6.5 {\AA}) were required
to reproduce all XANES and pre-edge features \cite{Ramos-EPL11}.
The spectral shape of the XMCD signal at the Mn K edge in our samples
(figure 1) changes from negative to positive at the edge (\textit{b,
b}'), it shows a resonance (\textit{c}) stemming from multiple scattering
of magnetic nearest neighbors, and finally a further peak (\textit{d})
is observed, associated with simultaneous excitation of 3p electrons
\cite{SubiasPRB1997}. In the sample with the largest XMCD signal
we also identify a resonance (\textit{p}) in the pre-edge range. These
spectral features are retained for all samples with only changes in
the amplitudes. The XMCD amplitude is defined here by the peak to
peak value (\textit{b, b'}), relative to the edge jump. This amplitude
increases as a function of the number of holes (figure
1, upper inset). All features are well reproduced by calculations
(figure 1, lower inset), except for the peak \textit{d},
confirming the multielectronic character of this structure.
ù
\begin{figure}
\includegraphics[bb=20bp 380bp 550bp 780bp,clip,scale=1]{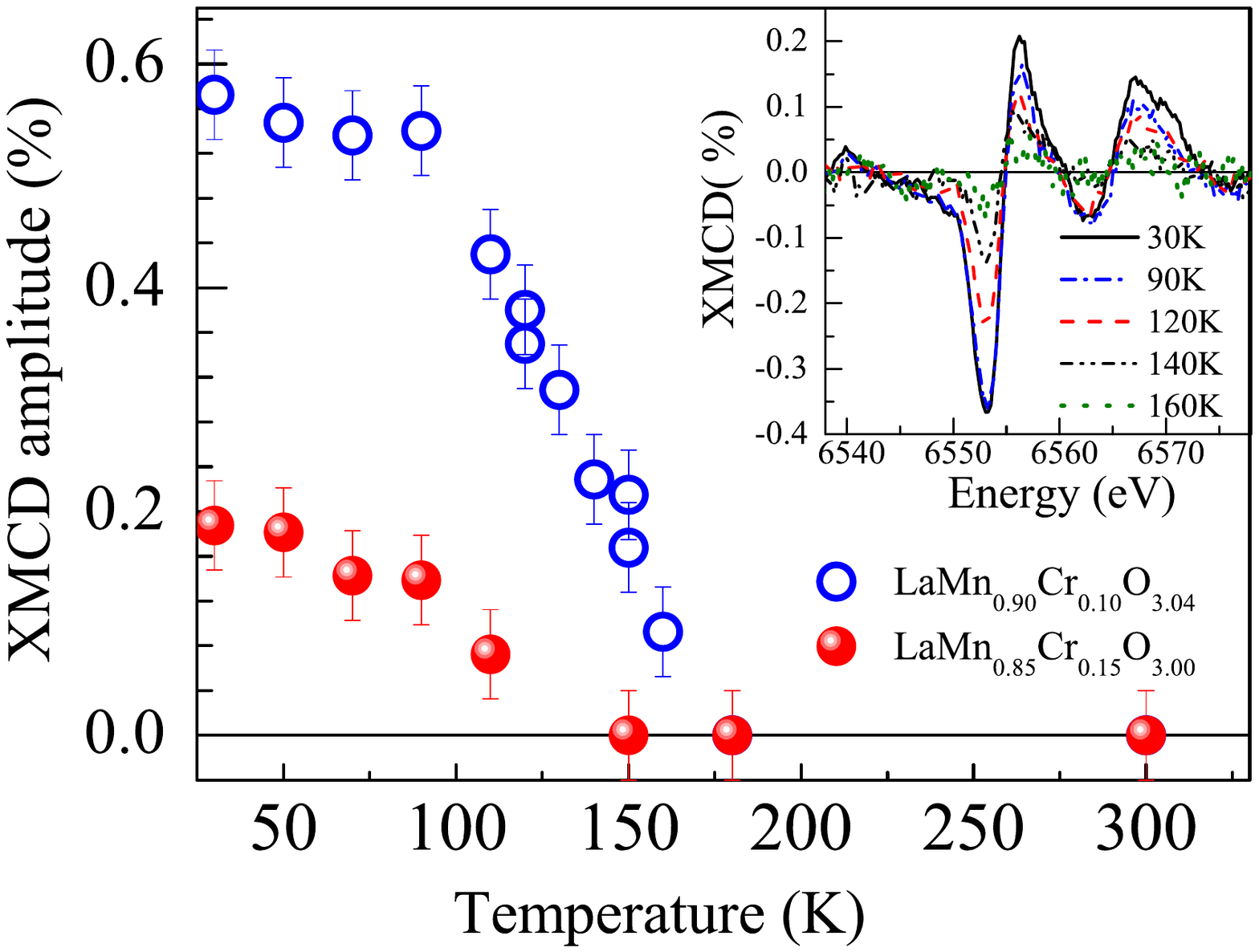}\caption{Temperature dependence (1.1 Tesla) of the Mn K edge XMCD amplitude
(see text) for $LaMn_{0.9}Cr_{0.1}O_{3.04}$ (open circles) and $LaMn_{0.85}Cr_{0.15}O_{3.00}$
(full circles). Inset XMCD signal of $LaMn_{0.9}Cr_{0.1}O_{3.04}$
at selected temperatures. }
\end{figure}

As a first step, to relate the XMCD signal to the macroscopic magnetic
properties \cite{Morales2004JAC,Morales2006PhysB}, we studied the
temperature dependence of the Mn K edge XMCD for two samples $LaMn_{0.9}Cr_{0.10}O_{3.04}$
and $LaMn_{0.95}Cr_{0.15}O_{3.00}$. This dependence (figure 2) agrees
with the ferromagnetic thermal evolution reported in these samples
\cite{Morales2004JAC,Morales2006PhysB}. We observe, as in macroscopic
measurements, a small reduction in the critical temperature in $LaMn_{0.95}Cr_{0.15}O_{3.00}$
with respect to $LaMn_{0.9}Cr_{0.10}O_{3.04}$ . At 30 K and 1.1 Tesla,
the XMCD amplitude ratio between the two samples, ie the ratio between
magnetic moment strength at the Mn site, is about the ratio of the
ferromagnetic component of the magnetic moment found by neutron scattering.

\begin{figure}
\includegraphics[bb=50bp 45bp 720bp 380bp,clip,scale=1]{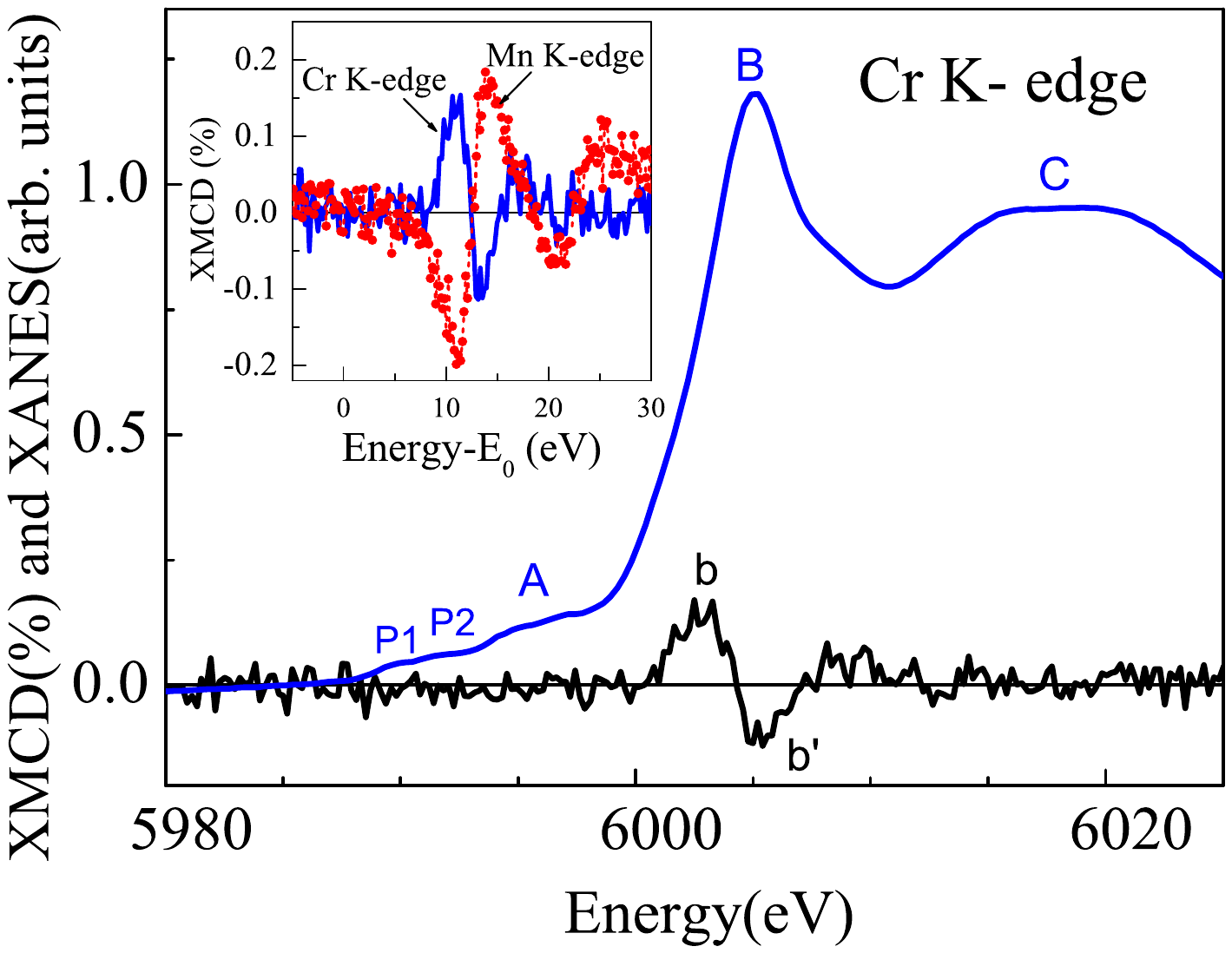}
\caption{XANES and XMCD (30 K, 3 Testla) at the Cr K edge for $LaMn_{0.9}Cr_{0.1}O_{3.00}$.
Inset : XMCD signals for $LaMn_{0.9}Cr_{0.1}O_{3.00}$ at the Cr K
edge and Mn K edge. The two energy scales have been shifted to coincide
at the main inflection point. }
\end{figure}

As concerns the Cr K edge (figure 3 for the sample $LaMn_{0.9}Cr_{0.1}O_{3.00}$),
the XMCD signal is opposite to that found at Mn K edge (figure 3,
inset), suggesting an antiferromagnetic Cr-Mn coupling. However, as
the K edge XMCD measures only the orbital part of the magnetic moment,
conclusions about the direction of the net moment should be supported
by calculations. The calculation at the Mn edge is made with a net
positive magnetic moment (number of spins up > number of spins down).
This gives the correct shape and sign of the XMCD signal (figure 1,
lower inset). At the Cr K edge, the calculations
have been performed both for a net positive (number of spins up >
number of spins down) and a net negative magnetic moment (number of
spins down> number of spins up). The two calculations give the same
shape but opposite signals. The Cr-XMCD signal that matches the experimental
result is the one with a net negative magnetic moment. So, we can
conclude that the net magnetic moments of Mn and Cr are opposite.
The XMCD spectral shape and sign are retained for all samples with
only changes in the amplitude. The coupling has the same antiferromagnetic
character regardless if $Mn^{4+}$ ions are present in the samples.
This demonstrates that this coupling is essentially determined by
$Mn^{3+}-Cr^{3+}$ interactions. 

\begin{figure}
\includegraphics[bb=100bp 50bp 700bp 360bp,clip,scale=1]{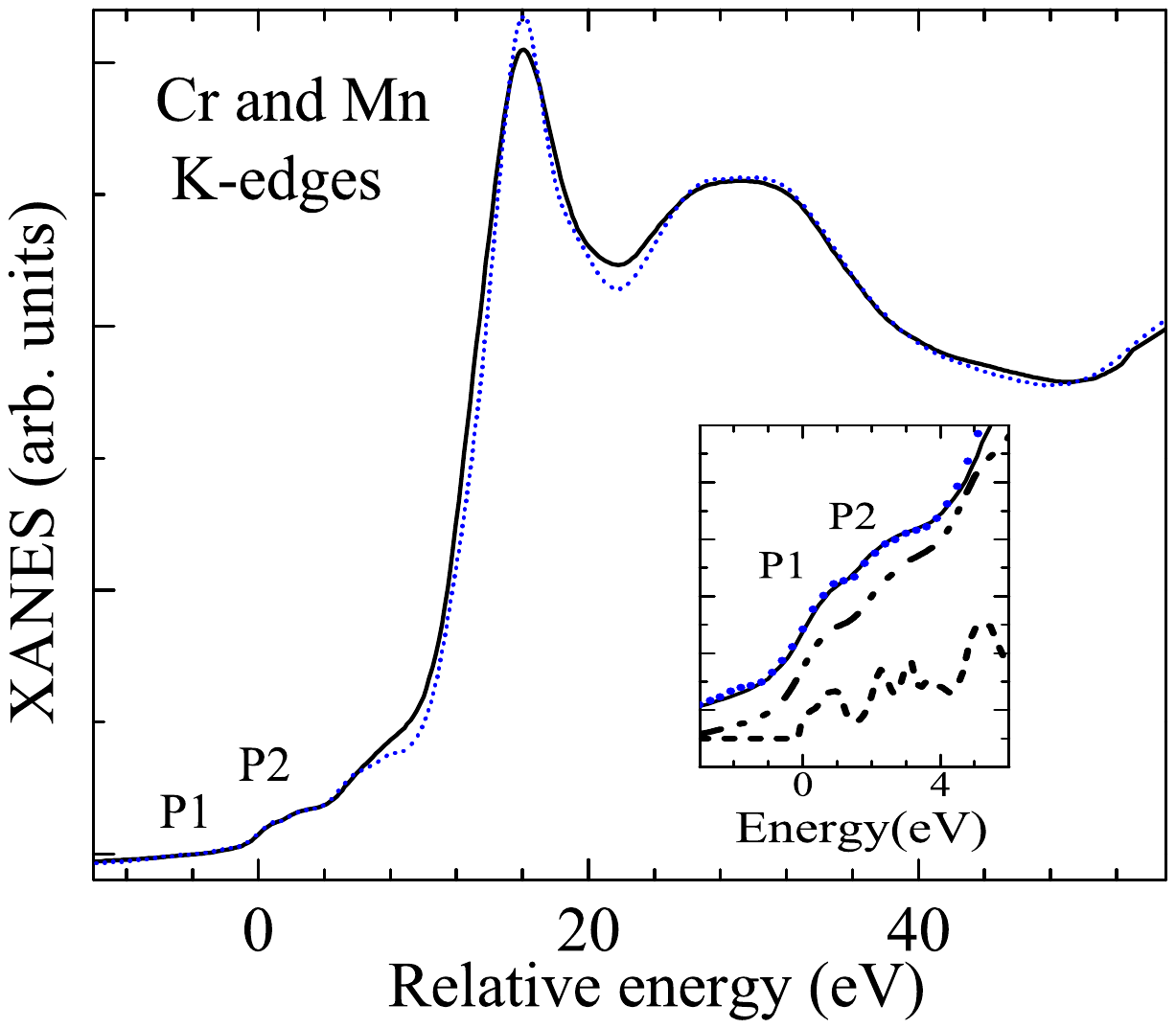}
\caption{XANES spectra at the Cr (dots) and Mn (plain line) K edges for the
sample $LaCr_{0.1}Mn_{0.9}O_{3.0}$. The energy scales have been shifted
to coincide at the first inflexion point. Inset: zoom in the pre-edge
range: the dashed lines correspond to\textit{ ab initio }calculations
before and after convolution. }
\end{figure}

We turn now to the characterization of the $Cr^{3+}$ local environment.
The EXAFS analysis of the Mn coordination shell gives the same parameters
for $LaMnO_{3}$ and for all $Mn^{4+}$ -free (x
= 0.05, 0.10 and 0.15) samples. Similar results
have been reported in Ga-substituted samples \cite{SanchezPhysRevB2004},
where a decrease of the JT distortion at the Mn sites is only observed
by EXAFS analysis  for x$\geqslant$0.2. However, due to the high
initial distortion of the $Mn^{3+}$ sites in $LaMnO_{3}$, the EXAFS
analysis in the low substituted range (x < 0.15)
has a limited sensitivity to small relaxations of the local distortion
and this result is not conclusive. More convincing is the striking
similarity between the XANES collected for the same sample at the
Cr and Mn K edges (figure 4), especially in pre-edge range (figure
4, inset). This indicates that the $Cr^{3+}$ ions in substitution
in the $LaMnO_{3}$ structure, adopt the same distorted local environment
as the $Mn^{3+}$ ions with negligible relaxation of the oxygen positions
around them. This outcome is corroborated by the comparison (figure
5) of the Mn K edge XANES in $LaMnO_{3}$ to the XANES for two samples
where 10\% of $e_{g}$ holes are introduced either by $Cr^{3+}$ substitution
($LaCr_{0.1}Mn_{0.9}O_{3.0}$) or by oxygen overstoichiometry ($LaMnO_{3.05}$).
In $LaMnO_{3.05}$ the pre-edge structure intensity is enhanced (figure
5-a) and the edge-jump is shifted towards higher
energy due to the presence of $Mn^{4+}$ ions (figure
5-b). An additional structure also emerges around 90 eV above the
edge (figure 5-c). Such structure is due to multiple scattering and
corresponds to an increase of the Mn-O-Mn angle, associated with the
rupture of the orbital order \cite{Souza-PRB04,Ramos-EPL11}. These
changes are the result of the symmetrization of the Mn environment
when doping with the introduction of 10\% of $e_{g}$ holes ($Mn^{4+}$)
by oxygen overstoichiometry. On the contrary, the introduction of
10\% of $e_{g}$ holes by $Cr^{3+}$ substitution does not change
the XANES spectrum, neither in the pre-edge and edge-jump
ranges (figure 5-a,b), nor around 90 eV above the edge (figure 5-c),
demonstrating that the Mn environment remains unchanged by Cr substitution.
Unlike $Mn^{4+}$ , $Cr^{3+}$ substitution does not significanly
alters the local site symmetry or the Mn-O-Mn superexchange
angle. We conclude that the JT is not significantly weakened by substitution
of $Mn^{3+}$ by $Cr^{3+}$ up to 15\%. This agrees
with the results obtained in Ga-substituted manganites for x=0.1\cite{SanchezPhysRevB2004}
. Nevertheless, while the nonmagnetic $Ga^{3+}$ ion
takes no part in the magnetic interactions, this is not the case of
the magnetically active $Cr^{3+}$ ion. It is worth
emphasizing here that $Cr^{3+}$ substitutes $Mn^{3+}$,
within the same local distorted environment , and that even through
it has an important role in the magnetic interactions. 

\begin{figure}
\includegraphics[bb=115bp 45bp 680bp 350bp,clip,scale=1]{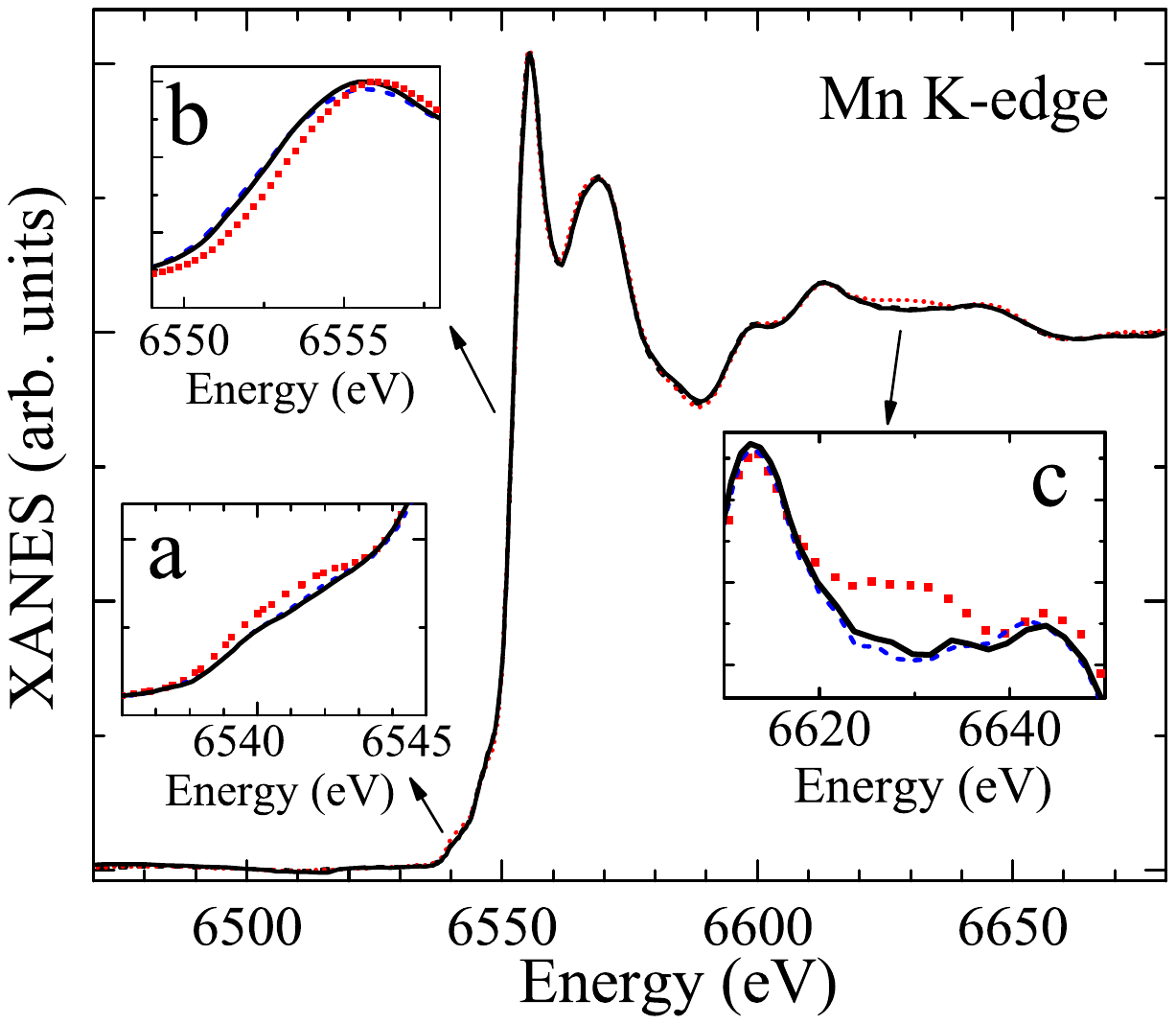}
\caption{XANES spectra at the Mn K edge in $LaMnO_{3}$ (black-plain line),
$LaCr_{0.1}Mn_{0.9}O_{3.0}$(blue-dashed line) and $LaMnO_{3.05}$(red-dots).
The insets a, b and c highlight the domains with spectral differences. }
\end{figure}

In low $x$~$LaMn_{1-x}Cr_{x}O_{3}$ compounds the $Cr^{3+}-Cr^{3+}$
interactions can be neglected in a first approximation. In these $Mn^{4+}$-free
samples the spin configuration stems only from the competition of
$Mn^{3+}-Cr^{3+}$ and $Mn^{3+}-Mn{}^{3+}$ interactions . We should
remind that in the parent compound A-AFM $LaMnO_{3}$, $Mn^{3+}-Mn{}^{3+}$
interactions are of two different natures. The $Mn{}^{3+}$ spins
are coupled ferromagnetically within ab-planes and and these planes
are coupled antiferromagnetically along the c-axis (sketched in figure
5). Our experimental results can be understood within the scenario
where the A- AFM order of $LaMnO_{3}$ is maintained, but with a small
FM component along the c-axis. When $Cr^{3+}$replaces one $Mn{}^{3+}$
in the network it will also experience both type of interactions.
Experimentally, we have shown that the signs of net magnetic moment
of $Mn{}^{3+}$ and $Cr^{3+}$ ions are opposite. Moreover, the $Cr^{3+}$
ions are substituted to $Mn{}^{3+}$ ions at exactly the same sites,
with no shift of oxygen towards $Cr^{3+}$. The Mn site distortion,
$e_{g}$ orbital occupancy and spin order remain essentially unmodified,
with an increase of the FM component along the c axis. As the $Cr^{3+}$
ion has no $e_{g}$ electron, hoping from the neighbouring ions- and
DE mechanism- becomes possible. Within the robust A-AFM scenario,
this results in a largely frustrated spin configuration. In a possible
configuration ferromagnetic DE between Cr and Mn magnetic moments
is effective within the ab-plane, but with an antiparallel alignment
of the magnetic component along the c-axis. Electron hoping along
the c-axis would be in this way less effective due to this antiparallel
alignment and electrical conductivity would be highly anisotropic.
With a Cr magnetic moment aligned antiparallel to the Mn spins along
the c axis, and parallel to the Mn spins in ab planes, the net FM
moment of Cr and Mn are then proportional and opposited along c axis.
The highly frustated magnetic arrangement is in agreement with the
observation of Morales and coworkers \cite{Morales2004JAC} who found,
for a same hole number, more magnetic frustration in $LaMn_{0.9}Cr_{0.1}O_{3}$
than in $LaMnO_{3.05}$. We should underline that the emergence of
FM component does not involve the same mechanism as A site substitution,
where the presence of $Mn^{4+}$ decreases the cooperative JT distortion,
thereby partially restoring the degenerescence of the $e_{g}$ orbital
and consequently weakening the orbital ordering. Our results rule
out the decrease of the JT distortion as the factor weakening the
antiferromagnetic $Mn^{3+}-Mn^{3+}$ superexchange. Orbital mixing
due to structural bias may likely play the important role. 
In conclusion, we studied the emergence of the FM component with the
inclusion of $Cr^{3+}$~ions in $LaMnO_{3}$. The net magnetic moment
of $Cr^{3+}$ ions is antiparallel to the moments of the $Mn^{3+}$
neighbours. Up to about 15\%, Cr to Mn substitution takes place without
rupture of the orbital ordering, the $Cr^{3+}$ ions occupying the
$Mn^{3+}$ sites without reduction of the local distortion. We ascertain
that the mechanism of emergence of 3D FM is different from the usual
A site doping, when $Mn^{4+}$ are formally introduced. Furthermore
our results underline the importance of local scale studies to safely
characterize the spin-spin interactions in strongly correlated systems. 

\begin{acknowledgments}
LNLS, SOLEIL and ESRF synchrotrons are acknowledged for beamtime.
\end{acknowledgments}


\begin{thebibliography}{34}
\expandafter\ifx\csname natexlab\endcsname\relax\def\natexlab#1{#1}\fi
\expandafter\ifx\csname bibnamefont\endcsname\relax
  \def\bibnamefont#1{#1}\fi
\expandafter\ifx\csname bibfnamefont\endcsname\relax
  \def\bibfnamefont#1{#1}\fi
\expandafter\ifx\csname citenamefont\endcsname\relax
  \def\citenamefont#1{#1}\fi
\expandafter\ifx\csname url\endcsname\relax
  \def\url#1{\texttt{#1}}\fi
\expandafter\ifx\csname urlprefix\endcsname\relax\def\urlprefix{URL }\fi
\providecommand{\bibinfo}[2]{#2}
\providecommand{\eprint}[2][]{\url{#2}}

\bibitem[{\citenamefont{Morales et~al.}(2005)\citenamefont{Morales, Allub,
  Alascio, Butera, and Caneiro}}]{Morales2005PRB72}
\bibinfo{author}{\bibfnamefont{L.}~\bibnamefont{Morales}},
  \bibinfo{author}{\bibfnamefont{R.}~\bibnamefont{Allub}},
  \bibinfo{author}{\bibfnamefont{B.}~\bibnamefont{Alascio}},
  \bibinfo{author}{\bibfnamefont{A.}~\bibnamefont{Butera}}, \bibnamefont{and}
  \bibinfo{author}{\bibfnamefont{A.}~\bibnamefont{Caneiro}},
  \bibinfo{journal}{Phys. Rev. B} \textbf{\bibinfo{volume}{72}},
  \bibinfo{pages}{132413} (\bibinfo{year}{2005}).

\bibitem[{\citenamefont{Bents}(1957)}]{BentsPRB1957}
\bibinfo{author}{\bibfnamefont{U.}~\bibnamefont{Bents}},
  \bibinfo{journal}{Phys. Rev.} \textbf{\bibinfo{volume}{106}},
  \bibinfo{pages}{255} (\bibinfo{year}{1957}).

\bibitem[{\citenamefont{Jonker}(1956)}]{JonkerPhysica56}
\bibinfo{author}{\bibfnamefont{G.}~\bibnamefont{Jonker}},
  \bibinfo{journal}{Physica} \textbf{\bibinfo{volume}{22}},
  \bibinfo{pages}{707} (\bibinfo{year}{1956}).

\bibitem[{\citenamefont{Ganguly et~al.}(2000)\citenamefont{Ganguly,
  Gopalakrishnan, and Yakhmi}}]{Ganguly_PhysicaB00}
\bibinfo{author}{\bibfnamefont{R.}~\bibnamefont{Ganguly}},
  \bibinfo{author}{\bibfnamefont{I.}~\bibnamefont{Gopalakrishnan}},
  \bibnamefont{and} \bibinfo{author}{\bibfnamefont{J.}~\bibnamefont{Yakhmi}},
  \bibinfo{journal}{Physica B: Condensed Matter}
  \textbf{\bibinfo{volume}{275}}, \bibinfo{pages}{308 } (\bibinfo{year}{2000}).

\bibitem[{\citenamefont{Gundakaram et~al.}(1996)\citenamefont{Gundakaram,
  Arulraj, Vanitha, Rao, Gayathri, Raychaudhuri, and
  Cheetham}}]{GundakaramJSSC1996}
\bibinfo{author}{\bibfnamefont{R.}~\bibnamefont{Gundakaram}},
  \bibinfo{author}{\bibfnamefont{A.}~\bibnamefont{Arulraj}},
  \bibinfo{author}{\bibfnamefont{P.}~\bibnamefont{Vanitha}},
  \bibinfo{author}{\bibfnamefont{C.~N.~R.} \bibnamefont{Rao}},
  \bibinfo{author}{\bibfnamefont{N.}~\bibnamefont{Gayathri}},
  \bibinfo{author}{\bibfnamefont{A.}~\bibnamefont{Raychaudhuri}},
  \bibnamefont{and} \bibinfo{author}{\bibfnamefont{A.~K.}
  \bibnamefont{Cheetham}}, \bibinfo{journal}{J. Solid State Chem.}
  \textbf{\bibinfo{volume}{127}}, \bibinfo{pages}{354} (\bibinfo{year}{1996}).

\bibitem[{\citenamefont{Qu et~al.}(2006)\citenamefont{Qu, Pi, Tan, Chen, Deng,
  and Zhang}}]{Qu_PRB06}
\bibinfo{author}{\bibfnamefont{Z.}~\bibnamefont{Qu}},
  \bibinfo{author}{\bibfnamefont{L.}~\bibnamefont{Pi}},
  \bibinfo{author}{\bibfnamefont{S.}~\bibnamefont{Tan}},
  \bibinfo{author}{\bibfnamefont{S.}~\bibnamefont{Chen}},
  \bibinfo{author}{\bibfnamefont{Z.}~\bibnamefont{Deng}}, \bibnamefont{and}
  \bibinfo{author}{\bibfnamefont{Y.}~\bibnamefont{Zhang}},
  \bibinfo{journal}{Phys. Rev. B} \textbf{\bibinfo{volume}{73}},
  \bibinfo{pages}{184407} (\bibinfo{year}{2006}).

\bibitem[{\citenamefont{Sun et~al.}(2001)\citenamefont{Sun, Tong, Xu, and
  Zhang}}]{Sun2001PRB63}
\bibinfo{author}{\bibfnamefont{Y.}~\bibnamefont{Sun}},
  \bibinfo{author}{\bibfnamefont{W.}~\bibnamefont{Tong}},
  \bibinfo{author}{\bibfnamefont{X.}~\bibnamefont{Xu}}, \bibnamefont{and}
  \bibinfo{author}{\bibfnamefont{Y.}~\bibnamefont{Zhang}},
  \bibinfo{journal}{Phys. Rev. B} \textbf{\bibinfo{volume}{63}},
  \bibinfo{pages}{174438} (\bibinfo{year}{2001}).

\bibitem[{\citenamefont{Farrell and Gehring}(2004)}]{Farell_NewJPhys04}
\bibinfo{author}{\bibfnamefont{J.}~\bibnamefont{Farrell}} \bibnamefont{and}
  \bibinfo{author}{\bibfnamefont{G.~A.} \bibnamefont{Gehring}},
  \bibinfo{journal}{New Journal of Physics} \textbf{\bibinfo{volume}{6}},
  \bibinfo{pages}{168} (\bibinfo{year}{2004}).

\bibitem[{\citenamefont{Cabeza et~al.}(1999)\citenamefont{Cabeza, Long,
  Severac, Bari, Muirhead, Francesconi, and Greaves}}]{Cabeza_JPCM99}
\bibinfo{author}{\bibfnamefont{O.}~\bibnamefont{Cabeza}},
  \bibinfo{author}{\bibfnamefont{M.}~\bibnamefont{Long}},
  \bibinfo{author}{\bibfnamefont{C.}~\bibnamefont{Severac}},
  \bibinfo{author}{\bibfnamefont{M.~A.} \bibnamefont{Bari}},
  \bibinfo{author}{\bibfnamefont{C.~M.} \bibnamefont{Muirhead}},
  \bibinfo{author}{\bibfnamefont{M.~G.} \bibnamefont{Francesconi}},
  \bibnamefont{and} \bibinfo{author}{\bibfnamefont{C.}~\bibnamefont{Greaves}},
  \bibinfo{journal}{J. Phys. Condens. Matter} \textbf{\bibinfo{volume}{11}},
  \bibinfo{pages}{2569} (\bibinfo{year}{1999}).

\bibitem[{\citenamefont{Barnabe et~al.}(1997)\citenamefont{Barnabe, Maignan,
  Hervieu, Damay, Martin, and Raveau}}]{Barnabe_APL97}
\bibinfo{author}{\bibfnamefont{A.}~\bibnamefont{Barnabe}},
  \bibinfo{author}{\bibfnamefont{A.}~\bibnamefont{Maignan}},
  \bibinfo{author}{\bibfnamefont{M.}~\bibnamefont{Hervieu}},
  \bibinfo{author}{\bibfnamefont{F.}~\bibnamefont{Damay}},
  \bibinfo{author}{\bibfnamefont{C.}~\bibnamefont{Martin}}, \bibnamefont{and}
  \bibinfo{author}{\bibfnamefont{B.}~\bibnamefont{Raveau}},
  \bibinfo{journal}{Appl. Phys. Lett.} \textbf{\bibinfo{volume}{71}},
  \bibinfo{pages}{3907} (\bibinfo{year}{1997}).

\bibitem[{\citenamefont{Capogna et~al.}(2008)\citenamefont{Capogna, Martinelli,
  Francesconi, Radaelli, Rodriguez~Carvajal, Cabeza, Ferretti, Castellano,
  Corridoni, and Pompeo}}]{CapognaPRB08}
\bibinfo{author}{\bibfnamefont{L.}~\bibnamefont{Capogna}},
  \bibinfo{author}{\bibfnamefont{A.}~\bibnamefont{Martinelli}},
  \bibinfo{author}{\bibfnamefont{M.~G.} \bibnamefont{Francesconi}},
  \bibinfo{author}{\bibfnamefont{P.~G.} \bibnamefont{Radaelli}},
  \bibinfo{author}{\bibfnamefont{J.}~\bibnamefont{Rodriguez~Carvajal}},
  \bibinfo{author}{\bibfnamefont{O.}~\bibnamefont{Cabeza}},
  \bibinfo{author}{\bibfnamefont{M.}~\bibnamefont{Ferretti}},
  \bibinfo{author}{\bibfnamefont{C.}~\bibnamefont{Castellano}},
  \bibinfo{author}{\bibfnamefont{T.}~\bibnamefont{Corridoni}},
  \bibnamefont{and} \bibinfo{author}{\bibfnamefont{N.}~\bibnamefont{Pompeo}},
  \bibinfo{journal}{Phys. Rev. B} \textbf{\bibinfo{volume}{77}},
  \bibinfo{pages}{104438} (\bibinfo{year}{2008}).

\bibitem[{\citenamefont{Ono et~al.}(2001)\citenamefont{Ono, Nakazono, and
  Oshima}}]{Ono_JMMM01}
\bibinfo{author}{\bibfnamefont{K.}~\bibnamefont{Ono}},
  \bibinfo{author}{\bibfnamefont{S.}~\bibnamefont{Nakazono}}, \bibnamefont{and}
  \bibinfo{author}{\bibfnamefont{M.}~\bibnamefont{Oshima}},
  \bibinfo{journal}{J. Magn. Magn. Mater.} \textbf{\bibinfo{volume}{226-230}},
  \bibinfo{pages}{869 } (\bibinfo{year}{2001}).

\bibitem[{\citenamefont{Toulemonde et~al.}(1998)\citenamefont{Toulemonde,
  Studer, Barnabe, Maignan, Martin, and Raveau}}]{Toulemonde_EPL98}
\bibinfo{author}{\bibfnamefont{O.}~\bibnamefont{Toulemonde}},
  \bibinfo{author}{\bibfnamefont{F.}~\bibnamefont{Studer}},
  \bibinfo{author}{\bibfnamefont{A.}~\bibnamefont{Barnabe}},
  \bibinfo{author}{\bibfnamefont{A.}~\bibnamefont{Maignan}},
  \bibinfo{author}{\bibfnamefont{C.}~\bibnamefont{Martin}}, \bibnamefont{and}
  \bibinfo{author}{\bibfnamefont{B.}~\bibnamefont{Raveau}},
  \bibinfo{journal}{Eur. Phys. J. B} \textbf{\bibinfo{volume}{4}},
  \bibinfo{pages}{159} (\bibinfo{year}{1998}).

\bibitem[{\citenamefont{Terashita et~al.}(2012)\citenamefont{Terashita, Cezar,
  Ardito, Bufaical, and Granado}}]{Terashita-PRB12}
\bibinfo{author}{\bibfnamefont{H.}~\bibnamefont{Terashita}},
  \bibinfo{author}{\bibfnamefont{J.~C.} \bibnamefont{Cezar}},
  \bibinfo{author}{\bibfnamefont{F.~M.} \bibnamefont{Ardito}},
  \bibinfo{author}{\bibfnamefont{L.~F.} \bibnamefont{Bufaical}},
  \bibnamefont{and} \bibinfo{author}{\bibfnamefont{E.}~\bibnamefont{Granado}},
  \bibinfo{journal}{Phys. Rev. B} \textbf{\bibinfo{volume}{85}},
  \bibinfo{pages}{104401} (\bibinfo{year}{2012}).

\bibitem[{\citenamefont{Cezar et~al.}(2010)\citenamefont{Cezar, Souza-Neto,
  Piamonteze, Tamura, Garcia, Carvalho, Neueschwander, Ramos, Tolentino,
  Caneiro et~al.}}]{Cezar-JSR10}
\bibinfo{author}{\bibfnamefont{J.~C.} \bibnamefont{Cezar}},
  \bibinfo{author}{\bibfnamefont{N.~M.} \bibnamefont{Souza-Neto}},
  \bibinfo{author}{\bibfnamefont{C.}~\bibnamefont{Piamonteze}},
  \bibinfo{author}{\bibfnamefont{E.}~\bibnamefont{Tamura}},
  \bibinfo{author}{\bibfnamefont{F.}~\bibnamefont{Garcia}},
  \bibinfo{author}{\bibfnamefont{E.~J.} \bibnamefont{Carvalho}},
  \bibinfo{author}{\bibfnamefont{R.~T.} \bibnamefont{Neueschwander}},
  \bibinfo{author}{\bibfnamefont{A.~Y.} \bibnamefont{Ramos}},
  \bibinfo{author}{\bibfnamefont{H.~C.~N.} \bibnamefont{Tolentino}},
  \bibinfo{author}{\bibfnamefont{A.}~\bibnamefont{Caneiro}},
  \bibnamefont{et~al.}, \bibinfo{journal}{J. Synchr. Rad.}
  \textbf{\bibinfo{volume}{17}}, \bibinfo{pages}{93} (\bibinfo{year}{2010}).

\bibitem[{\citenamefont{Deisenhofer et~al.}(2002)\citenamefont{Deisenhofer,
  Paraskevopoulos, Krug~von Nidda, and Loidl}}]{DeisenhoferPRB66}
\bibinfo{author}{\bibfnamefont{J.}~\bibnamefont{Deisenhofer}},
  \bibinfo{author}{\bibfnamefont{M.}~\bibnamefont{Paraskevopoulos}},
  \bibinfo{author}{\bibfnamefont{H.-A.} \bibnamefont{Krug~von Nidda}},
  \bibnamefont{and} \bibinfo{author}{\bibfnamefont{A.}~\bibnamefont{Loidl}},
  \bibinfo{journal}{Phys. Rev. B} \textbf{\bibinfo{volume}{66}},
  \bibinfo{pages}{054414} (\bibinfo{year}{2002}).

\bibitem[{\citenamefont{Zhou and Goodenough}(2008)}]{Zhou_PRB08}
\bibinfo{author}{\bibfnamefont{J.-S.} \bibnamefont{Zhou}} \bibnamefont{and}
  \bibinfo{author}{\bibfnamefont{J.~B.} \bibnamefont{Goodenough}},
  \bibinfo{journal}{Phys. Rev. B} \textbf{\bibinfo{volume}{77}},
  \bibinfo{pages}{172409} (\bibinfo{year}{2008}).

\bibitem[{\citenamefont{Mitchell et~al.}(1996)\citenamefont{Mitchell, Argyriou,
  Potter, Hinks, Jorgensen, and Bader}}]{Mitchell_PRB96}
\bibinfo{author}{\bibfnamefont{J.~F.} \bibnamefont{Mitchell}},
  \bibinfo{author}{\bibfnamefont{D.~N.} \bibnamefont{Argyriou}},
  \bibinfo{author}{\bibfnamefont{C.~D.} \bibnamefont{Potter}},
  \bibinfo{author}{\bibfnamefont{D.~G.} \bibnamefont{Hinks}},
  \bibinfo{author}{\bibfnamefont{J.~D.} \bibnamefont{Jorgensen}},
  \bibnamefont{and} \bibinfo{author}{\bibfnamefont{S.~D.} \bibnamefont{Bader}},
  \bibinfo{journal}{Phys. Rev. B} \textbf{\bibinfo{volume}{54}},
  \bibinfo{pages}{6172} (\bibinfo{year}{1996}).

\bibitem[{\citenamefont{Souza et~al.}(2004)\citenamefont{Souza, Souza-Neto,
  Ramos, Tolentino, and Granado}}]{Souza-PRB04}
\bibinfo{author}{\bibfnamefont{R.~A.} \bibnamefont{Souza}},
  \bibinfo{author}{\bibfnamefont{N.~M.} \bibnamefont{Souza-Neto}},
  \bibinfo{author}{\bibfnamefont{A.~Y.} \bibnamefont{Ramos}},
  \bibinfo{author}{\bibfnamefont{H.~C.~N.} \bibnamefont{Tolentino}},
  \bibnamefont{and} \bibinfo{author}{\bibfnamefont{E.}~\bibnamefont{Granado}},
  \bibinfo{journal}{Phys. Rev. B} \textbf{\bibinfo{volume}{70}},
  \bibinfo{pages}{214426} (\bibinfo{year}{2004}).

\bibitem[{\citenamefont{Monesi et~al.}(2005)\citenamefont{Monesi, Meneghini,
  Bardelli, Benfatto, Mobilio, Manju, and Sarma}}]{MonesiPRB05}
\bibinfo{author}{\bibfnamefont{C.}~\bibnamefont{Monesi}},
  \bibinfo{author}{\bibfnamefont{C.}~\bibnamefont{Meneghini}},
  \bibinfo{author}{\bibfnamefont{F.}~\bibnamefont{Bardelli}},
  \bibinfo{author}{\bibfnamefont{M.}~\bibnamefont{Benfatto}},
  \bibinfo{author}{\bibfnamefont{S.}~\bibnamefont{Mobilio}},
  \bibinfo{author}{\bibfnamefont{U.}~\bibnamefont{Manju}}, \bibnamefont{and}
  \bibinfo{author}{\bibfnamefont{D.~D.} \bibnamefont{Sarma}},
  \bibinfo{journal}{Phys. Rev. B} \textbf{\bibinfo{volume}{72}},
  \bibinfo{pages}{174104} (\bibinfo{year}{2005}).

\bibitem[{\citenamefont{Ramos et~al.}(2011)\citenamefont{Ramos, Souza-Neto,
  Tolentino, Bunau, Joly, Grenier, Iti\'e, Flank, Lagarde, and
  Caneiro}}]{Ramos-EPL11}
\bibinfo{author}{\bibfnamefont{A.~Y.} \bibnamefont{Ramos}},
  \bibinfo{author}{\bibfnamefont{N.~M.} \bibnamefont{Souza-Neto}},
  \bibinfo{author}{\bibfnamefont{H.~C.~N.} \bibnamefont{Tolentino}},
  \bibinfo{author}{\bibfnamefont{O.}~\bibnamefont{Bunau}},
  \bibinfo{author}{\bibfnamefont{Y.}~\bibnamefont{Joly}},
  \bibinfo{author}{\bibfnamefont{S.}~\bibnamefont{Grenier}},
  \bibinfo{author}{\bibfnamefont{J.-P.} \bibnamefont{Iti\'e}},
  \bibinfo{author}{\bibfnamefont{A.-M.} \bibnamefont{Flank}},
  \bibinfo{author}{\bibfnamefont{P.}~\bibnamefont{Lagarde}}, \bibnamefont{and}
  \bibinfo{author}{\bibfnamefont{A.}~\bibnamefont{Caneiro}},
  \bibinfo{journal}{EPL} \textbf{\bibinfo{volume}{96}}, \bibinfo{pages}{36002}
  (\bibinfo{year}{2011}).

\bibitem[{\citenamefont{Morales et~al.}(2004)\citenamefont{Morales,
  Rodriguez-Carvajal, and Caneiro}}]{Morales2004JAC}
\bibinfo{author}{\bibfnamefont{L.}~\bibnamefont{Morales}},
  \bibinfo{author}{\bibfnamefont{J.}~\bibnamefont{Rodriguez-Carvajal}},
  \bibnamefont{and} \bibinfo{author}{\bibfnamefont{A.}~\bibnamefont{Caneiro}},
  \bibinfo{journal}{J. Alloy Compd} \textbf{\bibinfo{volume}{369}},
  \bibinfo{pages}{97} (\bibinfo{year}{2004}).

\bibitem[{\citenamefont{Morales et~al.}(2006)\citenamefont{Morales, Caneiro,
  and James}}]{Morales2006PhysB}
\bibinfo{author}{\bibfnamefont{L.}~\bibnamefont{Morales}},
  \bibinfo{author}{\bibfnamefont{A.}~\bibnamefont{Caneiro}}, \bibnamefont{and}
  \bibinfo{author}{\bibfnamefont{M.}~\bibnamefont{James}},
  \bibinfo{journal}{Physica B} \textbf{\bibinfo{volume}{385}},
  \bibinfo{pages}{415} (\bibinfo{year}{2006}).

\bibitem[{\citenamefont{Morales et~al.}(2008)\citenamefont{Morales, Zysler, and
  Caneiro}}]{Morales2008JSSC}
\bibinfo{author}{\bibfnamefont{L.}~\bibnamefont{Morales}},
  \bibinfo{author}{\bibfnamefont{R.}~\bibnamefont{Zysler}}, \bibnamefont{and}
  \bibinfo{author}{\bibfnamefont{A.}~\bibnamefont{Caneiro}},
  \bibinfo{journal}{J. Solid State Chem.} \textbf{\bibinfo{volume}{181}},
  \bibinfo{pages}{1824} (\bibinfo{year}{2008}).

\bibitem[{\citenamefont{Morales and Caneiro}(2003)}]{Morales2003JSSC}
\bibinfo{author}{\bibfnamefont{L.}~\bibnamefont{Morales}} \bibnamefont{and}
  \bibinfo{author}{\bibfnamefont{A.}~\bibnamefont{Caneiro}},
  \bibinfo{journal}{J. Solid State Chem.} \textbf{\bibinfo{volume}{170}},
  \bibinfo{pages}{401} (\bibinfo{year}{2003}).

\bibitem[{\citenamefont{Tolentino et~al.}(2001)\citenamefont{Tolentino, Ramos,
  Alves, Barrea, Tamura, Cezar, and Watanabe}}]{Tolentino-JSR01}
\bibinfo{author}{\bibfnamefont{H.~C.~N.} \bibnamefont{Tolentino}},
  \bibinfo{author}{\bibfnamefont{A.~Y.} \bibnamefont{Ramos}},
  \bibinfo{author}{\bibfnamefont{M.~C.~M.} \bibnamefont{Alves}},
  \bibinfo{author}{\bibfnamefont{R.~A.} \bibnamefont{Barrea}},
  \bibinfo{author}{\bibfnamefont{E.}~\bibnamefont{Tamura}},
  \bibinfo{author}{\bibfnamefont{J.~C.} \bibnamefont{Cezar}}, \bibnamefont{and}
  \bibinfo{author}{\bibfnamefont{N.}~\bibnamefont{Watanabe}},
  \bibinfo{journal}{J. Synchrotron Rad.} \textbf{\bibinfo{volume}{8}},
  \bibinfo{pages}{1040} (\bibinfo{year}{2001}).

\bibitem[{\citenamefont{Proux et~al.}(2006)\citenamefont{Proux, Nassif, Prat,
  Ulrich, Lahera, Biquard, Menthonnex, and Hazemann}}]{Proux-JSR06}
\bibinfo{author}{\bibfnamefont{O.}~\bibnamefont{Proux}},
  \bibinfo{author}{\bibfnamefont{V.}~\bibnamefont{Nassif}},
  \bibinfo{author}{\bibfnamefont{A.}~\bibnamefont{Prat}},
  \bibinfo{author}{\bibfnamefont{O.}~\bibnamefont{Ulrich}},
  \bibinfo{author}{\bibfnamefont{E.}~\bibnamefont{Lahera}},
  \bibinfo{author}{\bibfnamefont{X.}~\bibnamefont{Biquard}},
  \bibinfo{author}{\bibfnamefont{J.}~\bibnamefont{Menthonnex}},
  \bibnamefont{and} \bibinfo{author}{\bibfnamefont{J.}~\bibnamefont{Hazemann}},
  \bibinfo{journal}{J. Synchr. Rad.} \textbf{\bibinfo{volume}{13}},
  \bibinfo{pages}{59} (\bibinfo{year}{2006}).

\bibitem[{\citenamefont{Baudelet et~al.}(2011)\citenamefont{Baudelet, Kong,
  Nataf, Cafun, Congeduti, Monza, Chagnot, and Iti\'e}}]{BaudeletHPR11}
\bibinfo{author}{\bibfnamefont{F.}~\bibnamefont{Baudelet}},
  \bibinfo{author}{\bibfnamefont{Q.}~\bibnamefont{Kong}},
  \bibinfo{author}{\bibfnamefont{L.}~\bibnamefont{Nataf}},
  \bibinfo{author}{\bibfnamefont{J.~D.} \bibnamefont{Cafun}},
  \bibinfo{author}{\bibfnamefont{A.}~\bibnamefont{Congeduti}},
  \bibinfo{author}{\bibfnamefont{A.}~\bibnamefont{Monza}},
  \bibinfo{author}{\bibfnamefont{S.}~\bibnamefont{Chagnot}}, \bibnamefont{and}
  \bibinfo{author}{\bibfnamefont{J.~P.} \bibnamefont{Iti\'e}},
  \bibinfo{journal}{High Pressure Research} \textbf{\bibinfo{volume}{31}},
  \bibinfo{pages}{136} (\bibinfo{year}{2011}).

\bibitem[{\citenamefont{Bridges et~al.}(2000)\citenamefont{Bridges, Booth,
  Kwei, Neumeier, and Sawatzky}}]{Bridges-PRB00}
\bibinfo{author}{\bibfnamefont{F.}~\bibnamefont{Bridges}},
  \bibinfo{author}{\bibfnamefont{C.~H.} \bibnamefont{Booth}},
  \bibinfo{author}{\bibfnamefont{G.~H.} \bibnamefont{Kwei}},
  \bibinfo{author}{\bibfnamefont{J.~J.} \bibnamefont{Neumeier}},
  \bibnamefont{and} \bibinfo{author}{\bibfnamefont{G.~A.}
  \bibnamefont{Sawatzky}}, \bibinfo{journal}{Phys. Rev. B}
  \textbf{\bibinfo{volume}{61}}, \bibinfo{pages}{R9237} (\bibinfo{year}{2000}).

\bibitem[{\citenamefont{Igarashi and Hirai}(1994)}]{Igarashi-PRB94}
\bibinfo{author}{\bibfnamefont{J.~I.} \bibnamefont{Igarashi}} \bibnamefont{and}
  \bibinfo{author}{\bibfnamefont{K.}~\bibnamefont{Hirai}},
  \bibinfo{journal}{Phys. Rev. B} \textbf{\bibinfo{volume}{50}},
  \bibinfo{pages}{17820} (\bibinfo{year}{1994}).

\bibitem[{\citenamefont{Guo}(1998)}]{Guo-PRB98}
\bibinfo{author}{\bibfnamefont{G.~Y.} \bibnamefont{Guo}},
  \bibinfo{journal}{Phys. Rev. B} \textbf{\bibinfo{volume}{57}},
  \bibinfo{pages}{10295} (\bibinfo{year}{1998}).

\bibitem[{\citenamefont{Bun{\u{a}}u and Joly}(2009)}]{Bunau-JPCM09}
\bibinfo{author}{\bibfnamefont{O.}~\bibnamefont{Bun{\u{a}}u}} \bibnamefont{and}
  \bibinfo{author}{\bibfnamefont{Y.}~\bibnamefont{Joly}}, \bibinfo{journal}{J.
  Phys.: Condens. Matter} \textbf{\bibinfo{volume}{21}},
  \bibinfo{pages}{345501} (\bibinfo{year}{2009}).

\bibitem[{\citenamefont{Subias et~al.}(1997)\citenamefont{Subias, Garcia,
  Proietti, and Blasco}}]{SubiasPRB1997}
\bibinfo{author}{\bibfnamefont{G.}~\bibnamefont{Subias}},
  \bibinfo{author}{\bibfnamefont{J.}~\bibnamefont{Garcia}},
  \bibinfo{author}{\bibfnamefont{M.~G.} \bibnamefont{Proietti}},
  \bibnamefont{and} \bibinfo{author}{\bibfnamefont{J.}~\bibnamefont{Blasco}},
  \bibinfo{journal}{Phys. Rev. B} \textbf{\bibinfo{volume}{56}},
  \bibinfo{pages}{8183} (\bibinfo{year}{1997}).

\bibitem[{\citenamefont{Sanchez et~al.}(2004)\citenamefont{Sanchez, Subias,
  Garcia, and Blasco}}]{SanchezPhysRevB2004}
\bibinfo{author}{\bibfnamefont{M.~C.} \bibnamefont{Sanchez}},
  \bibinfo{author}{\bibfnamefont{G.}~\bibnamefont{Subias}},
  \bibinfo{author}{\bibfnamefont{J.}~\bibnamefont{Garcia}}, \bibnamefont{and}
  \bibinfo{author}{\bibfnamefont{J.}~\bibnamefont{Blasco}},
  \bibinfo{journal}{Phys. Rev. B} \textbf{\bibinfo{volume}{69}},
  \bibinfo{pages}{184415} (\bibinfo{year}{2004}).

\end{thebibliography}
\end{document}